\documentstyle[12pt,epsf]{article}
 
\input{axodraw.sty}

\def\firsttext{
Michael Spira \\
{\it CERN Theory Division, CH-1211 Geneva 23, Switzerland}

\end{center}

\vspace*{9cm}

\begin{abstract}
The determination of the full SUSY QCD corrections to the
production of squarks and gluinos at hadron colliders is reviewed. The
inclusion of the NLO corrections stabilizes the theoretical predictions of
the various production cross sections significantly and leads to sizeable
enhancements of the most relevant cross sections. We discuss the
phenomenological consequences of the results on present and future
experimental analyses. Finally we investigate the impact of the corrections
on the transverse momentum and rapidity distributions.
\end{abstract}
}

\def\moriond{
\pagestyle{empty}
\begin{document}

\begin{titlepage}

\vspace*{0.8cm}

\begin{center}
{\large \sc Squark and Gluino Production at Hadron Colliders}
\\~\\~\\

\firsttext

\end{titlepage}
}

\def\cern{
\begin{document}

\begin{titlepage}

\begin{flushright}
CERN-TH/97-110 \\
hep-ph/9705403
\end{flushright}
\vspace*{0.2cm}

\vspace*{0.8cm}

\renewcommand{\thefootnote}{\fnsymbol{footnote}}

\begin{center}
{\large \sc Squark and Gluino Production at Hadron Colliders\footnote{Invited
talk given at XXXIInd Rencontre de Moriond: QCD and High Energy Hadronic
Interactions, Les Arcs, France, March 22--29, 1997.}}
\\~\\~\\

\firsttext

\vspace*{\fill}

\begin{flushleft}
CERN-TH/97-110 \\
hep-ph/9705403 \\
May 1997
\end{flushleft}

\end{titlepage}
}

\def\firstpage{\cern}

\catcode`@=11

\def\@citex[#1]#2{\if@filesw\immediate\write\@auxout
        {\string\citation{#2}}\fi
\def\@citea{}\@cite{\@for\@citeb:=#2\do
        {\@citea\def\@citea{,}\@ifundefined
        {b@\@citeb}{{\bf ?}\@warning
        {Citation `\@citeb' on page \thepage \space undefined}}
        {\csname b@\@citeb\endcsname}}}{#1}}

\newif\if@cghi
\def\cite{\@cghitrue\@ifnextchar [{\@tempswatrue
        \@citex}{\@tempswafalse\@citex[]}}
\def\citelow{\@cghifalse\@ifnextchar [{\@tempswatrue
        \@citex}{\@tempswafalse\@citex[]}}
\def\@cite#1#2{{$\!^{#1)}$\if@tempswa\typeout
        {IJCGA warning: optional citation argument
        ignored: `#2'} \fi}}
\newcommand{\citeup}{\cite}


\def\citer{\@ifnextchar [{\@tempswatrue\@citexr}{\@tempswafalse\@citexr[]}}

%

\def\@citexr[#1]#2{\if@filesw\immediate\write\@auxout
        {\string\citation{#2}}\fi
\def\@citea{}\@cite{\@for\@citeb:=#2\do
        {\@citea\def\@citea{--}\@ifundefined
        {b@\@citeb}{{\bf ?}\@warning
        {Citation `\@citeb' on page \thepage \space undefined}}
        {\csname b@\@citeb\endcsname}}}{#1}}
\catcode`@=12

\newcommand{\lsim}{\raisebox{-0.13cm}{~\shortstack{$<$ \\[-0.07cm] $\sim$}}~}
\newcommand{\gsim}{\raisebox{-0.13cm}{~\shortstack{$>$ \\[-0.07cm] $\sim$}}~}

\newcommand{\sq}{\tilde{q}}
\newcommand{\sqb}{\bar{\tilde{q}}}
\newcommand{\gl}{\tilde{g}}
\newcommand{\MSSM}{\mbox{${\cal MSSM}$}}
\newcommand{\SUSY}{\mbox{${\cal SUSY}$}}
\newcommand{\TeV}{\mbox{Te$\!$V}}
\newcommand{\GeV}{\mbox{Ge$\!$V}}
\newcommand{\tb}{\mbox{tg$\beta$}}

\newcommand{\nn}{\noindent}

 \newcommand{\zp}[3]{{Z.\ Phys.} {\bf #1} (19#2) #3}
 \newcommand{\np}[3]{{Nucl.\ Phys.} {\bf #1} (19#2)~#3}
 \newcommand{\pl}[3]{{Phys.\ Lett.} {\bf #1} (19#2) #3}
 \newcommand{\pr}[3]{{Phys.\ Rev.} {\bf #1} (19#2) #3}
 \newcommand{\prl}[3]{{Phys.\ Rev. Lett.} {\bf #1} (19#2) #3}
 \newcommand{\prep}[3]{{\sl Phys. Rep.} {\bf #1} (19#2) #3}
 \newcommand{\fp}[3]{{\sl Fortschr. Phys.} {\bf #1} (19#2) #3}
 \newcommand{\nc}[3]{{\sl Nuovo Cimento} {\bf #1} (19#2) #3}
 \newcommand{\ijmp}[3]{{\sl Int. J. Mod. Phys.} {\bf #1} (19#2) #3}
 \newcommand{\ptp}[3]{{\sl Prog. Theo. Phys.} {\bf #1} (19#2) #3}
 \newcommand{\sjnp}[3]{{\sl Sov. J. Nucl. Phys.} {\bf #1} (19#2) #3}
 \newcommand{\cpc}[3]{{\sl Comp. Phys. Commun.} {\bf #1} (19#2) #3}
 \newcommand{\mpl}[3]{{\sl Mod. Phys. Lett.} {\bf #1} (19#2) #3}
 \newcommand{\cmp}[3]{{\sl Commun. Math. Phys.} {\bf #1} (19#2) #3}
 \newcommand{\jmp}[3]{{\sl J. Math. Phys.} {\bf #1} (19#2) #3}
 \newcommand{\nim}[3]{{\sl Nucl. Instr. Meth.} {\bf #1} (19#2) #3}
 \newcommand{\el}[3]{{\sl Europhysics Letters} {\bf #1} (19#2) #3}
 \newcommand{\ap}[3]{{\sl Ann. of Phys.} {\bf #1} (19#2) #3}
 \newcommand{\jetp}[3]{{\sl JETP} {\bf #1} (19#2) #3}
 \newcommand{\jetpl}[3]{{\sl JETP Lett.} {\bf #1} (19#2) #3}
 \newcommand{\acpp}[3]{{\sl Acta Physica Polonica} {\bf #1} (19#2) #3}
 \newcommand{\vj}[4]{{\sl #1~}{\bf #2} (19#3) #4}
 \newcommand{\ej}[3]{{\bf #1} (19#2) #3}
 \newcommand{\vjs}[2]{{\sl #1~}{\bf #2}}
 \newcommand{\hep}[1]{{hep--ph/}{#1}}
 \newcommand{\desy}[1]{{DESY-Report~}{#1}}


\firstpage

\renewcommand{\baselinestretch}{1.5}
\normalsize

\section{Introduction}
The search for Higgs bosons and supersymmetric particles is among the most
important endeavors of present and future high energy physics. The novel
colored particles,
squarks and gluinos, can be searched for at the Tevatron, a $p\bar p$ collider
with a c.m.\ energy of 1.8 TeV, and the future LHC, a $pp$ collider with a
c.m. energy of 14 TeV. Until now the search at the Tevatron has set the most
stringent bounds on their masses. At the 95\% CL, gluinos have
to be heavier than about 175 GeV, while squarks with masses below about 175 GeV
have been excluded for gluino masses below $\sim 300$ GeV \cite{bounds}. In the
$R$-parity-conserving MSSM, supersymmetric particles can only be produced
in pairs. All supersymmetric particles will decay to the lightest
supersymmetric particle (LSP), which is most likely to be a neutralino, stable
thanks to conserved $R$-parity.
Thus the final signatures for the production of supersymmetric particles will
mainly be jets and missing transverse energy, which is carried away by the
invisible neutral LSP.

Squarks and gluinos can be produced via the processes
\begin{equation}
p\bar p/pp \to \sq \sq, \sq \sqb, \gl \gl, \sq \gl + X \, .
\label{eq:proc}
\end{equation}
In the following we consider the production of all squarks and gluinos except
stops. We assume the squarks to be mass degenerate, which is a reasonable
approximation for all squark flavors except stops, while the light quarks
($u,d,s,c,b$) will be treated as massless particles. The production of stop
pairs requires the inclusion of mass splitting and mixing effects and will be
investigated elsewhere \cite{stops}. In eq.~(\ref{eq:proc}) a summation over all
possible squark flavors and charge conjugate final states should be
implicitly understood. The calculation of the LO cross sections has been
performed a long time ago \cite{lo}. Since the [unphysical] scale dependence
of the LO quantities amounts to about 50\%, the determination of the NLO
corrections is necessary in order to gain a reliable theoretical prediction,
which can be used in present and future experimental analyses.

\section{SUSY QCD corrections}
The evaluation of the full SUSY QCD corrections splits into two pieces, the
virtual corrections, generated by virtual particle exchanges, and the
real corrections, which originate from gluon radiation and the corresponding
crossed processes with three-particle final states \cite{sqgl}.

\subsection{Virtual corrections}
\begin{figure}[hbt]
\begin{center}
\setlength{\unitlength}{1pt}
\begin{picture}(120,60)(20,20)

\Gluon(0,20)(50,20){-3}{5}
\Gluon(0,80)(50,80){3}{5}
\Gluon(50,50)(75,80){-3}{4}
\DashLine(100,20)(50,20){5}
\DashLine(50,80)(100,80){5}
\DashLine(50,20)(50,80){5}
\put(-15,78){$g$}
\put(-15,18){$g$}
\put(75,48){$g$}
\put(105,18){$\sqb$}
\put(105,78){$\sq$}

\end{picture}
\begin{picture}(120,60)(-40,20)

\Gluon(0,20)(50,20){-3}{5}
\Gluon(0,80)(50,80){-3}{5}
\Gluon(75,20)(75,80){-3}{5}
\ArrowLine(75,80)(75,20)
\ArrowLine(75,20)(50,20)
\ArrowLine(50,20)(50,80)
\ArrowLine(50,80)(75,80)
\DashLine(100,20)(75,20){5}
\DashLine(75,80)(100,80){5}
\put(-15,78){$g$}
\put(-15,18){$g$}
\put(40,48){$q$}
\put(85,48){$\gl$}
\put(105,18){$\sqb$}
\put(105,78){$\sq$}

\end{picture}  \\
\setlength{\unitlength}{1pt}
\renewcommand{\baselinestretch}{1}
\normalsize
\caption[]{\label{fg:virt} \it Typical diagrams of the virtual corrections.}
\renewcommand{\baselinestretch}{1.5}
\normalsize
\end{center}
\end{figure}
The one-loop virtual corrections are built up by gluon, gluino, quark and
squark exchange
contributions [see Fig.~\ref{fg:virt}]. They have to be contracted with the
LO matrix elements. The calculation of the one-loop contributions has been
performed in dimensional regularization, leading to the extraction of
ultraviolet, infrared and collinear singularities as poles in
$\epsilon = (4-n)/2$. For the chiral $\gamma_5$ coupling we have used the naive
scheme, which is well justified in the present analysis at the one-loop level.
We have explicitly checked that after summing all virtual corrections no
quadratic divergences are left over, in accordance with the general property
of supersymmetric theories. The renormalization has been performed by
identifying the squark and gluino masses with their pole masses, and defining
the strong
coupling in the $\overline{\rm MS}$ scheme including five light flavors in the
corresponding $\beta$ function. The massive particles, i.e.\ squarks, gluinos
and top quarks, have been decoupled by subtracting their contribution at
vanishing momentum transfer \cite{decouple}. In dimensional regularization,
there is a mismatch between the gluonic degrees of freedom [d.o.f. = $n-2$] and
those of the gluino [d.o.f. = $2$], so that SUSY is explicitly broken. In
order to restore SUSY in the physical observables in the massless limit, an
additional finite counter-term is required for the renormalization of the novel
$\sq \gl \bar q$ vertex \cite{count}.

\subsection{Real corrections}
\begin{figure}[hbt]
\begin{center}
\setlength{\unitlength}{1pt}
\begin{picture}(120,70)(20,20)

\Gluon(0,20)(50,50){-3}{5}
\Gluon(0,80)(50,50){3}{5}
\Gluon(25,65)(75,95){3}{5}
\DashLine(50,50)(100,80){5}
\DashLine(100,20)(50,50){5}
\put(-15,78){$g$}
\put(-15,18){$g$}
\put(80,93){$g$}
\put(105,18){$\sqb$}
\put(105,78){$\sq$}

\end{picture}
\begin{picture}(170,70)(-40,20)

\Gluon(0,20)(50,50){-3}{5}
\ArrowLine(0,80)(50,50)
\ArrowLine(50,50)(100,50)
\DashLine(100,50)(150,80){5}
\Gluon(100,50)(125,35){-3}{3}
\ArrowLine(100,50)(125,35)
\ArrowLine(125,35)(150,10)
\DashLine(125,35)(150,35){5}
\put(-15,78){$q$}
\put(-15,18){$g$}
\put(102,30){$\gl^*$}
\put(155,78){$\sq$}
\put(155,33){$\sqb$}
\put(155,8){$q$}

\end{picture}  \\
\setlength{\unitlength}{1pt}
\renewcommand{\baselinestretch}{1}
\normalsize
\caption[]{\label{fg:real} \it Typical diagrams of the real corrections.}
\renewcommand{\baselinestretch}{1.5}
\normalsize
\end{center}
\end{figure}
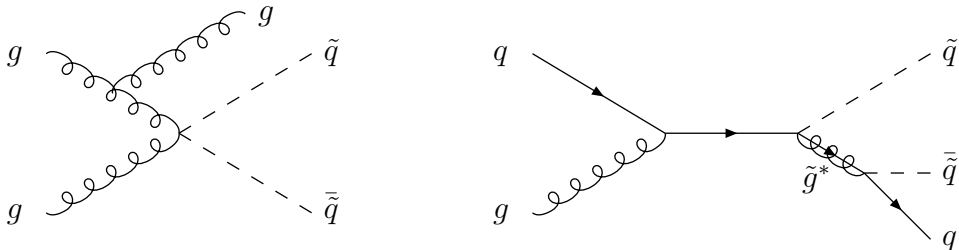
The real corrections originate from the radiation of a gluon in all possible
ways and from the crossed processes by interchanging the gluon of
the final state against a light quark in the initial state. The phase-space
integration of the final-state particles has been performed in $n=4-2\epsilon$
dimensions, leading to the extraction of infrared and collinear singularities
as poles in $\epsilon$. In order to isolate the singularities we have
introduced a cutoff $\Delta$ in the invariant mass of, say, the $\sqb g$ pair,
which separates soft and hard gluon radiation. After evaluating all
angular integrals and adding the virtual and real corrections, the infrared
singularities cancel. The left-over collinear singularities are universal and
are absorbed in the renormalization of the parton densities at NLO. We defined
the parton densities in the conventional $\overline{\rm MS}$ scheme including
five light flavors, i.e.\ the squark, gluino and top quark contributions are
not included in the mass factorization. Finally we end up with an ultraviolet,
infrared and collinear finite partonic cross section, which is independent
of the cutoff for $\Delta\to 0$.

However, there is an additional class of physical singularities, which have to
be regularized~\cite{sqgl}. In the second diagram of Fig.~\ref{fg:real} an
intermediate $\sq \gl^*$ state is produced, before the [off-shell] gluino splits
into a $q\sqb$ pair. If the gluino mass is larger than the common squark mass,
and the partonic c.m.\ energy is larger than the sum of the squark and gluino
masses, the intermediate gluino can be produced on its mass-shell. Thus the
real corrections to $\sq \sqb$ production contain a contribution of $\sq \gl$
production. The residue of this part corresponds to $\sq \gl$ production with
the subsequent gluino decay $\gl \to \sqb q$, which is already contained
in the LO cross section
of $\sq \gl$ pair production, including all final-state cascade decays. Thus
this term has to be subtracted in order to derive a well-defined production
cross section. Analogous subtractions emerge in all reactions: if the gluino
mass is larger than the squark mass, the contributions from $\gl \to \sq \bar
q, \sqb q$ have to be subtracted, and in the reverse case the contributions of
squark decays into gluinos have to subtracted.

\begin{figure}[b]
\vspace*{-1.5cm}

\hspace*{0.5cm}
\epsfxsize=15cm \epsfbox{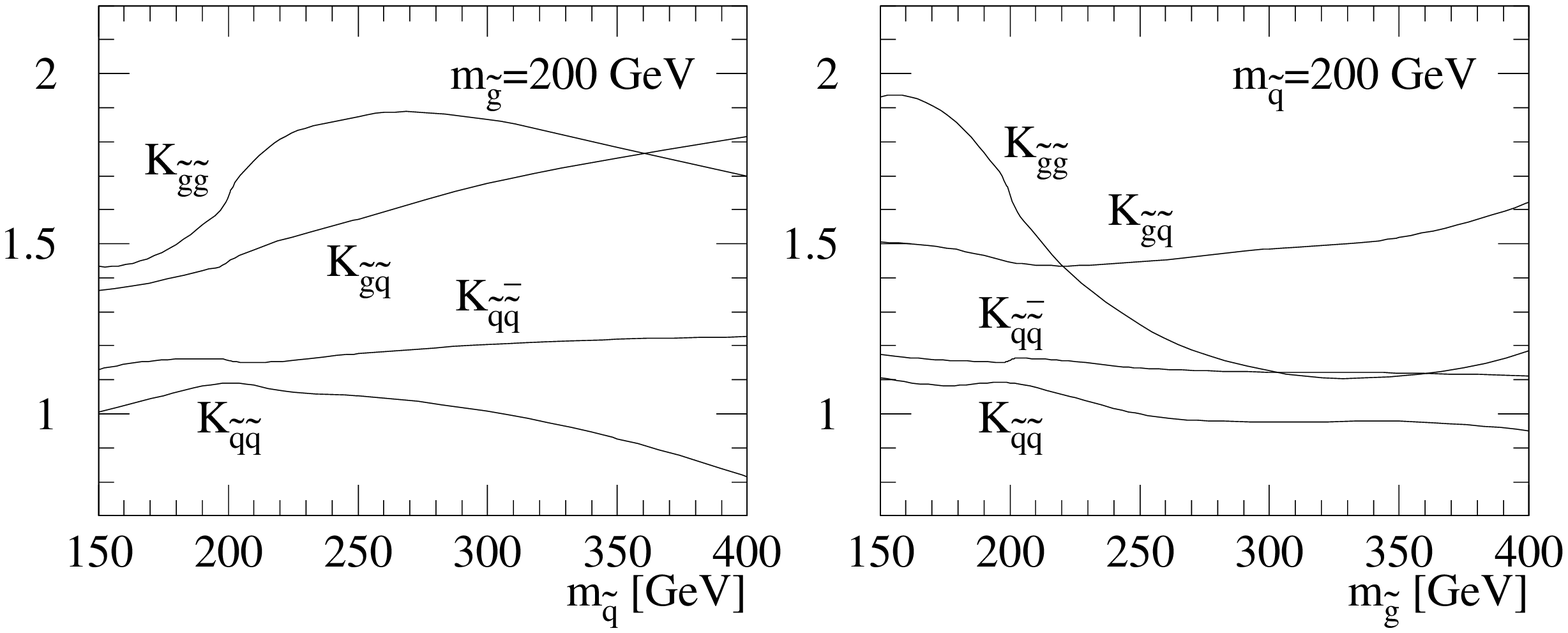}
\vspace*{-1.2cm}

\renewcommand{\baselinestretch}{1}
\normalsize
\caption[]{\label{fg:kfac} \it K factors of the different squark and gluino
production cross sections at the Tevatron. Parton density: GRV(94) with $Q=m$.
Top mass: $m_t=175$ GeV.}
\renewcommand{\baselinestretch}{1.5}
\normalsize
\end{figure}
\section{Results}
The hadronic cross sections can be obtained from the partonic ones by
convolution with the corresponding parton densities. We have performed the
numerical analysis for the Tevatron and the LHC. For the natural
renormalization/factorization scale choice $Q=m$, where $m$ denotes the
average mass of the final-state SUSY particles, the SUSY QCD corrections are
large and positive, increasing the total
cross sections by 10--90\% \cite{sqgl}. This is shown in Fig.~\ref{fg:kfac},
where the K factors, defined as the ratios of the NLO and LO cross sections,
are presented as a function of the corresponding SUSY particle mass for the
Tevatron.

\begin{figure}[t]
\vspace*{-8cm}

\hspace*{3cm}
\epsfxsize=13cm \epsfbox{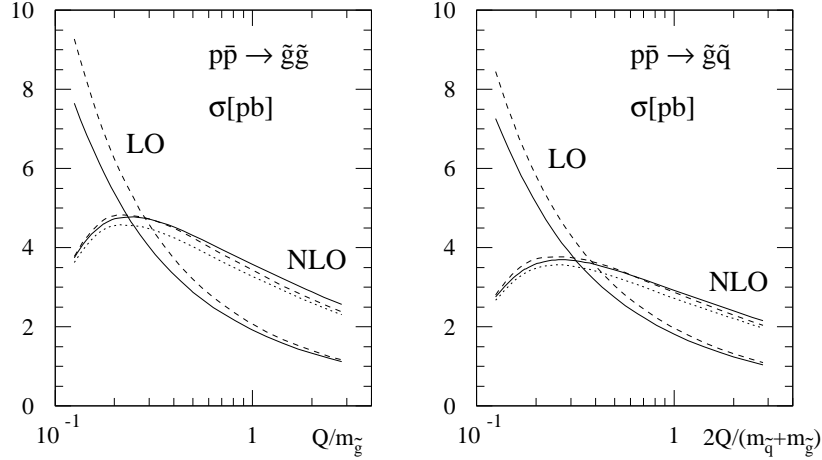}
\vspace*{-1.3cm}

\renewcommand{\baselinestretch}{1}
\normalsize
\caption[]{\label{fg:scale} \it Scale and parton density dependence of the
total $\gl \gl$ and $\gl \sq$ production cross sections at the Tevatron in LO
and NLO. Parton densities: GRV(94) (solid), CTEQ3 (dashed) and MRS(A')
(dotted); mass parameters: $m_{\sq}=280$ GeV, $m_{\gl}=200$ GeV and $m_t=175$
GeV.}
\renewcommand{\baselinestretch}{1.5}
\normalsize
\end{figure}
We have investigated the residual scale dependence in LO and NLO, which is
presented in Fig.~\ref{fg:scale}. The inclusion
of the NLO corrections reduces the LO scale dependence by a factor 3--4 and
reaches a typical level of $\sim 15\%$, which serves as an estimate of the
remaining theoretical uncertainty \cite{sqgl}. Moreover, the dependence on
different sets of parton densities is rather weak and leads to an additional
uncertainty of $\sim 10\%$ \cite{sqgl}. In order to quantify the effect of the
NLO corrections on the
search for squarks and gluinos at hadron colliders, we have extracted the
SUSY particle masses corresponding to several fixed values of the production
cross sections. These masses are increased by 10--30 GeV at the Tevatron and
by 10--50 GeV at the LHC, thus enhancing the present bounds on the squark and
gluino masses significantly \cite{sqgl}.

\begin{figure}[hbt]
\vspace*{-1.5cm}

\hspace*{0cm}
\epsfxsize=17cm \epsfbox{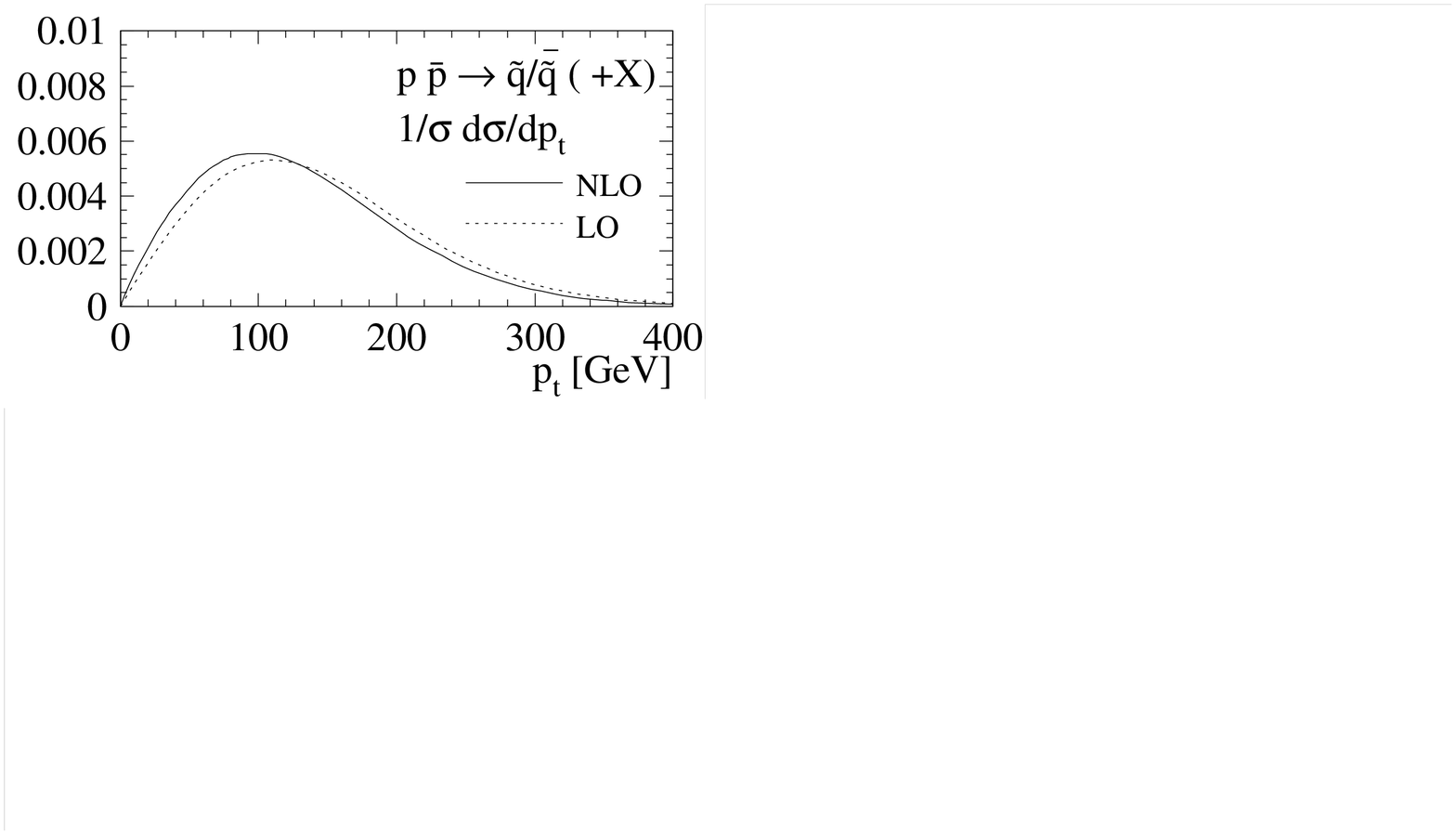}

\vspace*{-10.9cm}

\hspace*{8cm}
\epsfxsize=17cm \epsfbox{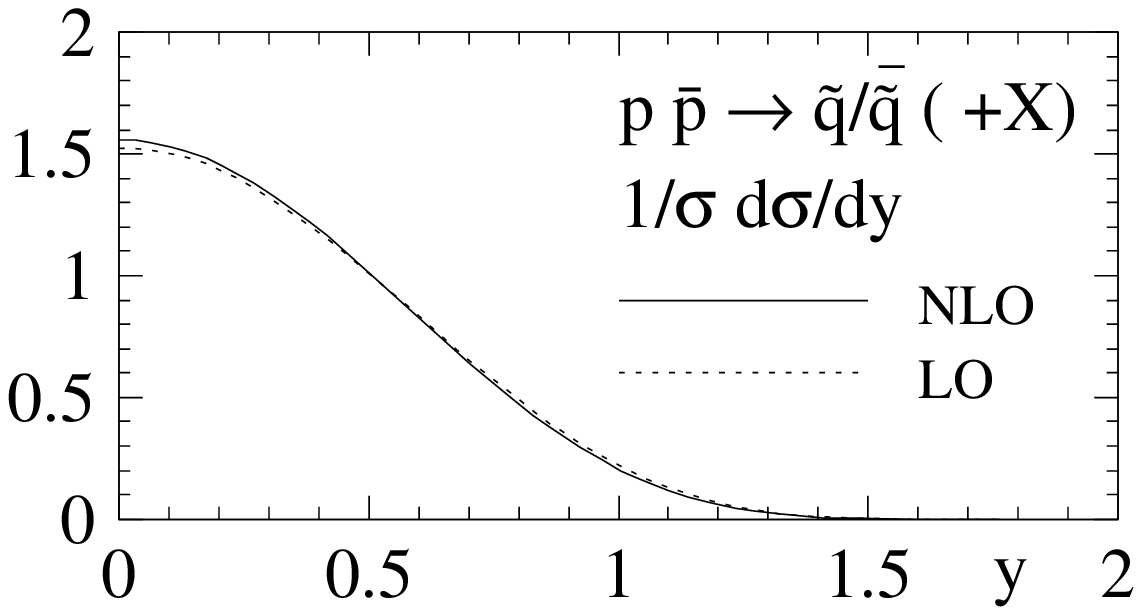}
\vspace*{-5.5cm}

\renewcommand{\baselinestretch}{1}
\normalsize
\caption[]{\label{fg:pty} \it Normalized transverse-momentum and rapidity
distributions of $p\bar p\to \sq \sqb + X$ at the Tevatron in LO (dotted)
and NLO (solid). Parton densities:
GRV('94) with $Q=m$; mass parameters: $m_{\sq}=280$ GeV, $m_{\gl}=200$ GeV
and $m_t=175$ GeV.}
\renewcommand{\baselinestretch}{1.5}
\normalsize
\end{figure}
Finally we have evaluated the QCD-corrected transverse-momentum and rapidity
distributions for all different processes. As can be inferred from
Fig.~\ref{fg:pty}, the modification of the normalized distributions in NLO
compared to LO is less than 10\% for the transverse-momentum
distributions and negligible for the rapidity distributions. Thus it
is a sufficient approximation to rescale the LO distributions uniformly by the
K factors of the total cross sections \cite{sqgl}. \\

\noindent {\bf Acknowledgements} \\
I would like to thank W.\ Beenakker, R.\ H\"opker and P.M.\ Zerwas for their
pleasant collaboration in the presented work.

\end{document}